\documentclass[conference]{IEEEtran}
\IEEEoverridecommandlockouts
\usepackage{cite,url}
\usepackage{amsmath,amssymb,amsfonts}
\usepackage{graphicx}
\usepackage{textcomp}
\usepackage{xcolor}
\usepackage{comment}
\usepackage{amsmath,amssymb} 
\usepackage{color}
\usepackage[noend]{algpseudocode}
\usepackage{algorithmicx,algorithm}
\usepackage{multirow}
\usepackage{subfigure}

\usepackage{bm}

\algblock{Input}{EndInput}
\algnotext{EndInput}
\algblock{Output}{EndOutput}
\algnotext{EndOutput}

\def\BibTeX{{\rm B\kern-.05em{\sc i\kern-.025em b}\kern-.08em
    T\kern-.1667em\lower.7ex\hbox{E}\kern-.125emX}}

\begin{document}

\title{Boost  CTR Prediction for New Advertisements\\ via Modeling Visual Content
}

\author{
\IEEEauthorblockN{Tan Yu,  Zhipeng Jin,  Jie Liu, Yi Yang, Hongliang Fei, Ping Li} \\
\IEEEauthorblockA{Cognitive Computing Lab, Baidu Research}
\IEEEauthorblockA{Baidu Search Ads (Phoenix Nest), Baidu Inc.}
10900 NE 8th St. Bellevue, Washington 98004, USA\\
No. 10 Xibeiwang East Road, Beijing 100193, China \\\\
\{tanyu01, jinzhipeng, liujie34,  yangyi15, hongliangfei, liping11\}@baidu.com
}

\maketitle

\begin{abstract}
Existing advertisements click-through rate (CTR) prediction models are mainly dependent on behavior ID features, which are learned based on the historical user-ad interactions.
Nevertheless, behavior ID features relying on historical user behaviors are not feasible to describe new ads without previous interactions with users. To overcome the limitations of behavior  ID features in modeling new ads, we exploit the visual content in ads to boost the performance of CTR prediction models. Specifically, we map each ad into a set of visual IDs based on its visual content. These visual IDs are further used for generating the visual embedding for enhancing CTR prediction models. We formulate the learning of visual IDs into a supervised quantization problem. Due to a lack of class labels for commercial images in advertisements, we exploit image textual descriptions as the supervision to optimize the image extractor for generating effective visual IDs. Meanwhile, since the hard quantization is non-differentiable, we soften the quantization operation to make it support the end-to-end network training. After mapping each image into visual IDs, we learn the embedding for each visual ID based on the historical user-ad interactions accumulated in the past. Since the visual ID embedding depends only on the visual content, it generalizes well to new ads. Meanwhile, the visual ID embedding complements the ad behavior ID embedding. Thus, it can considerably boost the performance of the CTR prediction models previously relying on behavior  ID features for both new ads and ads that have accumulated rich user behaviors.   After incorporating the visual ID embedding in the CTR prediction model of Baidu online advertising, the average CTR of ads improves by $1.46\%$, and the total charge increases by   $1.10\%$.
\end{abstract}


\section{Introduction}

Online advertising platforms serve personalized advertisements based on users' potential interests. For example, Baidu Search Ads (a.k.a. ``Phoenix Nest'') has been successfully using ultra-high dimensional input data and ultra-large-scale deep neural networks for training CTR (Click-Through Rate) models since 2013~\cite{fan2019mobius,zhao2019aibox,fei2021gemnn}.  In an advertising platform based on eCPM (effective Cost Per Mille), the advertisements fed to a specific user are ranked by the product of the bid price offered by the advertisers and the user's predicted CTR from the CTR prediction model. Intuitively, it tends to put the advertisements with high predicted CTR and bid price at the top of the rank list to attract potential customers for advertisers and achieve high profit for the online advertising platform.

Since the predicted CTR has a substantial impact on the rank of the displayed ads,  the deviation of the predicted CTR  from the actual CTR has a significant influence on the revenue of advertisers and the advertising platform. If the predicted CTR of an ad is lower than the actual CTR, the ad might not get exposed to customers. In this case,   the advertisers will not gain the expected revenues, and the advertising platform also loses the charges which it should have attained. On the other hand, if the predicted CTR of an ad is higher than the actual CTR, the ad will be misplaced at the top of the rank list but does not lead to the expected amount of clicks from customers. Then the advertisers will be overcharged by the advertising platform and cannot gain a reasonable profit. Thus,  an effective CTR prediction model is a critical component for an eCPM advertising platform to achieve satisfactory advertising effects for the advertisers and abundant revenues for the advertising platform.

With the prompt progress achieved in machine learning in the past decade, we have witnessed the rapid evolution in the architecture of the CTR prediction model. The earliest works are mainly based on linear logistic regression (LR) model~\cite{lee2012estimating}, non-linear gradient boosting decision trees (GBDT)~\cite{friedman2001greedy,li2007mcrank,zheng2007regression,li2010robust,he2014practical,li2022package}, Bayesian models~\cite{graepel2010web} or  factorization machines (FM)~\cite{rendle2010factorization,ta2015factorization,hu2018collaborative}. These models take a shallow structure and thus cannot effectively describe high-order latent patterns in the user-ad behavior. Inspired by the great success of deep learning in computer vision and natural language processing, researchers attempt to build deep neural networks (DNN) for CTR prediction. Factorization-machine supported neural network (FNN)~\cite{zhang2016deep} feeds the output of FM into a deep fully-connected~neural network. Convolutional Click Prediction Model (CCPM)~\cite{liu2015convolutional} predicts the CTR by a deep convolutional neural network.   Zhang~{et al.}~\cite{zhang2014sequential}  utilize the recurrent neural network to model the sequence of user behaviors for CTR prediction. The following works~\cite{cheng2016wide,qu2016product,guo2017deepfm,wang2017deep,wang2018dkn,lian2018xdeepfm,song2019autoint,zhou2019deep,huang2019fibinet,fan2019mobius,feng2019deep,ma2019hierarchical,zhao2019aibox,pi2020search,rendle2020neural,fei2021gemnn,xu2021agile}  explore more advanced neural networks for more effectively modeling higher-order patterns.  Although extensive efforts have been devoted to improving the CTR prediction on the model side,  the data side in the CTR prediction has been relatively less exploited. The lack of training data is commonly encountered for many machine learning tasks, including  CTR prediction model training.


The existing mainstream CTR prediction model is based on the behavior ID features based on the historical interactions between users and ads.  Each ad, as well as each user, is assigned a unique behavior ID.  Each behavior ID of a user/ad is mapped to a vector, termed as embedding vector. The ad embedding vectors and user embedding vectors are learned jointly based on the historical behaviors of users on ads.  Since the user/ad embedding vectors are learned based on historical behaviors,  the quality of the learned features is heavily dependent on the richness of users' past behaviors on ads.   When a new ad is added to the advertising system, we have no access to the user-ad behavior on this new ad. It leads to the lack of the training data issue, and the behavior ID embedding for this new ad is not reliable at all for predicting the CTR of any user on the ad. The lack of training data for new ads  is normally termed the cold-start problem.

  A straightforward solution to solving the cold-start problem is using content-based features by understanding the visual content of ads~\cite{mo2015image,ge2018image}.
  Specifically, Mo~\emph{et al.}~\cite{mo2015image}  learn  image features of  display ads directly from raw pixels  and user feedback in the target task. Ge~\emph{et al.}~\cite{ge2018image} exploits both ad image features  and user behavior image features. Nevertheless, due to the high computational cost of image  feature extraction,  Mo~\emph{et al.}~\cite{mo2015image} and Ge~\emph{et al.}~\cite{ge2018image}  adopt a pre-trained  image feature extractor to get the image feature in the offline phase and then use the extracted image feature for predicting CVR/CTR in the online phase.  Since the feature extractor is not related to CVR/CTR prediction in the training phase, the extracted image features might not be effective for CVR/CTR prediction. Zhao~\emph{et al.}~\cite{zhao2019you}  devise a pre-ranking model. When an advertiser uploads  ads, the offline pre-ranking model determines inferior and superior ads based on their visual content.  Only the superior ads with attractive visual content will be fed into the online ranker for further ad display. Since the pre-ranking model  is conducted in the offline phase,  the inefficiency caused by complex feature extraction is no longer an issue, and it makes the end-to-end training of feature extractor feasible. The offline pre-ranking model is trained  by the accumulated historical data from all users in a learn-to-rank manner.  Since the training data is collected from all users, it reveals the preference of the majority of users and might not reveal the preference of a specific user for personalized advertising.

\begin{figure}[h]
\centering
\includegraphics[width=3.4in]{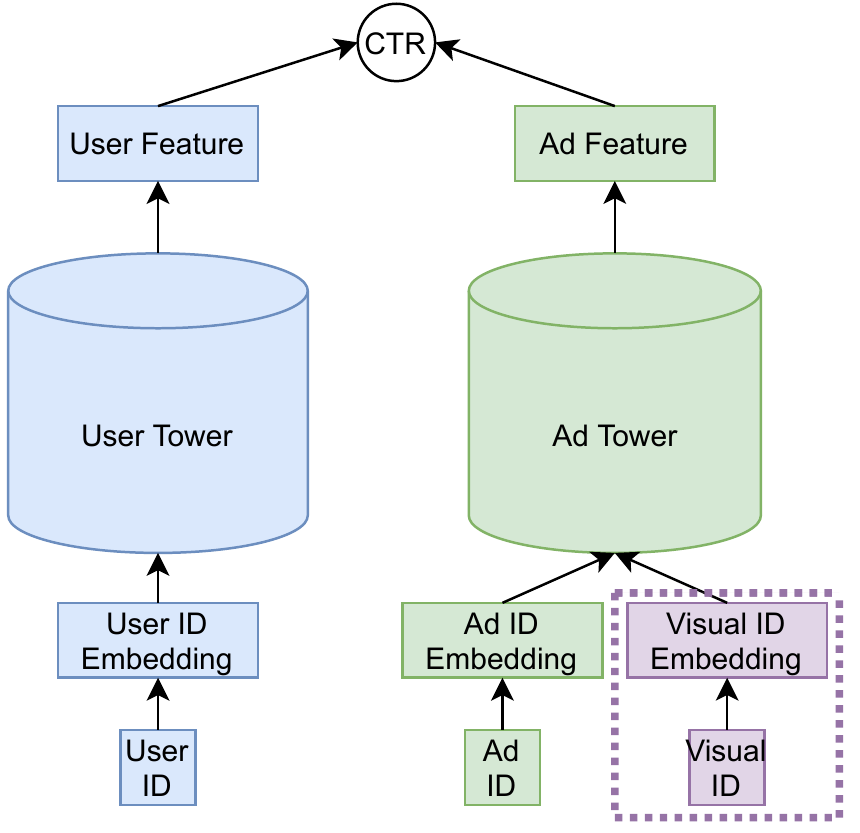}
\caption{The structure of the proposed model. Different from the traditional CTR prediction model using only ad behavior ID embedding and user behavior ID embedding,  our model additionally incorporates the visual content by devising a visual ID embedding for each ad (the purple rectangular).  Since the visual ID embedding is only dependent on the visual content of the ad, it generalizes well for the new ads.}
\label{structure}
\end{figure}

In this work, we seek to  model the visual content in ads for boosting the performance CTR prediction effectively.  We map the visual content of an ad into a set of visual indices (IDs) through supervised  clustering.  
 After we obtain the visual IDs for ads, we learn a visual embedding vector for each visual ID based on the user's clicks on ads. The learned visual ID embedding vector for  each ad can effectively describe the visual content  in an ad and can well generalize to  ads newly added to the advertising system. Meanwhile, the visual content encoded in the visual ID embedding vector is complementary to the ad behavior ID embedding vector, which is beneficial to CTR prediction for not only new ads  but also the ads which have accumulated rich user behaviors.
In this case, as visualized in Figure~\ref{structure},  both ad behavior ID embedding and visual ID embedding are the input of the CTR prediction model, and they work together to achieve a more accurate CTR prediction.

To generate the visual IDs for ads, a straightforward solution is using k-means to cluster the visual features from an off-the-shelf image feature extractor, \emph{e.g.}, a convolutional neural network.
Nevertheless, the image features  from the off-the-shelf image extractor such as ResNet-50~\cite{he2016deep} pre-trained on ImageNet~\cite{deng2009imagenet} might not be effective for discriminating the commercial images.  To bridge the domain gap between the pre-trained dataset and the commercial application, we can fine-tune the pre-trained image extractor on commercial images in ads.  Traditionally, fine-tuning the  ResNet-50 requires   human labours to annotate the label of each image, which is prohibitively expensive  for the huge-scale commercial image corpus. Inspired by the recent success achieved by CLIP~\cite{radford2021learning} in self-supervised learning,  we exploit the textual description of the image in  an ad as the supervision and construct a contrastive learning task to optimize the image feature extractor.  After contrastive learning through supervision from textual descriptions of images, we can obtain the visual features. But it still deviates from our goal of obtaining  the visual IDs.  Thus, we devise a supervised clustering module based on the learned codebooks when fine-tuning the image feature extractor. Since the hard assignment in clustering is not differentiable, we soften the cluster assignment  for achieving end-to-end training.
Meanwhile, to enrich the codebooks for partitioning  the visual feature space into finer cells, we adapt  residual quantization and  product quantization.
After training the image feature extractor as well as the codebooks for supervised clustering using the textual description through contrastive learning, we use them to generate the visual IDs.
The visual IDs are further mapped into visual ID embeddings, which will be the input of the CTR prediction model.




\vspace{0.05in}

   In summary, the \textbf{contributions} of this work are four-fold:
\begin{itemize}
\item We propose to incorporate  visual content in the  CTR prediction model through learning the visual ID embedding for ads. The learned visual ID embedding generalizes well to new ads and complements the ad behavior ID embedding.   \vspace{0.0in}

\item We devise an effective approach to learn the visual ID  using the supervision from textual descriptions in a contrast-learning manner. We formulate the visual ID learning into a supervised clustering problem based on learned codebooks. To support  end-to-end learning, we soften the cluster assignment and make it differentiable.\vspace{0.05in}

\item To enrich the codebooks for  partitioning the visual feature space into finer cells, we adopt both residual quantization and product quantization.  \vspace{0.05in}

\item We launched the proposed  CTR prediction model incorporating the visual content in Baidu's online advertising system. After launching, we achieved a $1.46\%$ improvement in CTR and a   $1.10\%$ increase~in~total~charges.


\end{itemize}

\section{Related Work}

\textbf{CTR prediction model.} CTR prediction is a long-standing problem in online advertising and recommender system.  The focus of CTR prediction is on learning an effective embedding for users as well as products (ads). Early works on embedding learning are mainly based  on  linear regression (LR)~\cite{jahrer2010combining,mcmahan2013ad,chapelle2014modeling,wang2021efficient} or factorization machine (FM)~\cite{koren2009matrix,rendle2010factorization,rendle2012factorization,juan2016field,hu2018collaborative}.
Specifically, LR-based methods use a linear projection to map the features into an embedding vector. The weights of the linear projection are optimized based on the binary cross-entropy loss. In parallel, FM-based methods  map features into a latent space and model the interactions between users and products (ads) through the inner product of their
embedding vectors. Inspired by the great success achieved by  deep learning in computer vision~\cite{he2016deep} and natural language processing~\cite{vaswani2017attention}, several methods
exploit deep neural networks to learn  embedding vectors.   Specifically, CNN-based CTR prediction model~\cite{liu2015convolutional} exploits the interactions between neighbors in the feature space for  enhancing the discriminating power of embedding features. Zhang \emph{et al.}~\cite{zhang2016deep}  utilizes deep neural networks to enhance the  features from traditional embedding methods such as  FMs, restricted Boltzmann machines (RBMs), and denoising auto-encoders (DAEs) for generating more effective embedding vectors. Wide$\&$Deep learning~\cite{cheng2016wide}   jointly trains wide linear models and deep neural networks to combine their benefits  for the recommender system. Product-based Neural Networks (PNN)~\cite{qu2016product} investigates the interactions between diffident features through inner-product and outer-product operations in neural networks for CTR prediction. Deep crossing~\cite{shan2016deep} builds a deep neural network that automatically combines features to produce superior models.   DeepFM~\cite{guo2017deepfm} combines a factorization machine (FM) and a deep neural network (DNN). In DeepFM, the FM extracts low-order features, whereas the DNN generates the high-order features. DKN~\cite{wang2018dkn} develops  a  word-entity-aligned and knowledge-aware convolutional neural network to fuse s semantic-level and knowledge-level representations of news.   Deep Interest Network (DIN)~\cite{zhou2018deep} models the user's rich historical behaviors through a sequence model. Deep Interest Evolution Network (DIEN)~\cite{zhou2019deep} improves DIN by using a more advanced sequence model, GRU. Deep Session Interest Network~(DSIN)~\cite{feng2019deep} also focuses on modeling users' behaviors in a sequence of sessions.  It adopts Transformer~\cite{vaswani2017attention}, which has shown better performance than GRU in modeling sentences and documents.  Search-based Interest Model
(SIM)~\cite{pi2020search} models the user's lifelong sequential behavior data through Transformer to exploit richer user behaviors. In these aforementioned methods, the CTR prediction is mainly based on the user's ID embedding and the item's ID embedding learned from the historical user behaviors. Thus, they might not perform well when the users' historical behaviors are not sufficient. Different from above works, we utilize the visual content to complement  the user's ID embedding and the item's ID embedding to achieve good CTR prediction performance for new ads and new users.


\vspace{0.1in}
\noindent \textbf{Visual features in CTR prediction.} Mo \emph{et al.}~\cite{mo2015image} feed the raw images in products (ads) into a convolutional neural network (CNN) and train the CNN through user-click supervision. It addresses the cold start problem when the ID features are not reliable when the historical behaviors are not rich enough. Ge \emph{et al.}~\cite{ge2018image}  extract image features not only on the products (ads) side but also on the user side. Zhao~\emph{et al.}~\cite{zhao2019you} observes the efficiency problem encountered in~\cite{mo2015image,ge2018image} and moves the image feature extractor to the offline phase.  Specifically, before ranking the products (ads) through the CTR prediction model, they pre-rank the products (ads) based on their visual features. Category-specific CNN (CSCNN)~\cite{liu2020category} early fuses the product category with the visual image when modeling the product visual content and meanwhile devise an efficient architecture to make the online deployment of CNN model feasible.
Wang~\emph{et al.}~\cite{wang2021hybrid} proposes a hybrid bandit model using visual content as priors for creative ranking. Different from the above-mentioned methods, we map each ad into a visual ID. We train our image encoder and the visual ID generator based on the text-image pairs in an end-to-end manner.  Then we learn the visual ID embedding using the  objective of optimizing CTR prediction accuracy. Thus, our visual ID embedding not only encodes the visual content  but also effectively describes the user-item interactions. Recently, the teams in Baidu Research and Baidu Search Ads developed a series of works~\cite{yu2020combo,yu2021multi,yu2022tree} to exploit the use of vision BERT for boosting the text-visual relevance for video ads.

\vspace{0.05in}
\section{Visual ID Generation}
\vspace{0.05in}

\textbf{Supervision signal.}
We denote the image in the ad by $I$  and denote the visual index of the image by $\mathrm{ID}_I$.  Straightforwardly, $\mathrm{ID}_I$ can be obtained from k-means clustering on ad image features extracted from an off-the-shelf image feature extractor, \emph{e.g.}, ResNet pre-trained on ImageNet dataset. Nevertheless, there is a domain gap between the natural images in ImageNet and the commercial images in the ads. A solution to adapting the target domain is fine-tuning the pre-trained ResNet on the commercial images in the ads. But there are no class labels available for supervising  the network fine-tuning, and it is prohibitively expensive to annotate the image labels manually. Inspired by the great success achieved by CLIP~\cite{radford2021learning} in self-supervised learning, we exploit the textual description of the images in the ad as the supervision to fine-tune the network through contrastive learning.  We formulate the process of learning visual IDs for images in ads into a deep supervised clustering problem. Below we introduce the details.




\vspace{0.08in}
\noindent \textbf{Deep supervised clustering.} We denote  the textual description of the image $I$ by $T$. We denote the image feature extractor by $\mathrm{Encoder}_{\mathrm{img}}$ and that for encoding the textual description by $\mathrm{Encoder}_{\mathrm{txt}}$.  We denote the image feature of $I$ generated from $\mathrm{Encoder}_{\mathrm{img}}$ by $\mathbf{x} \in \mathbb{R}^d$ and the text feature  of $T$ from  $\mathrm{Encoder}_{\mathrm{txt}}$ by $\mathbf{y} \in \mathbb{R}^d$.  That is,
\begin{equation}
\mathbf{x} = \mathrm{Encoder}_{\mathrm{img}}(I), \mathbf{y} = \mathrm{Encoder}_{\mathrm{txt}}(T).
\end{equation}
To make the clustering feasible,  we devise a dictionary  denoted  by $\mathcal{C} = \{\mathbf{c}_i\}_{i=1}^N$.  $\mathcal{C}$ are weights of the network which are randomly initialized and optimized through contrastive learning.
Straightforwardly, we can map the visual feature $\mathbf{x}$ from the image extractor to its closest codeword in $\mathcal{C}$.  In this case,  the visual ID of the image $I$  is just the index of its closest codeword:

\begin{equation}
\label{hard}
\mathrm{ID}_I = \mathrm{argmax}_{i\in[1,N]} -\|\mathbf{c}_i - \mathbf{x}\|_2,
\end{equation}
where $\|.\|_2$ denotes the $L_2$ norm. The visual feature  $\mathbf{i}$ is mapped to $\mathbf{c}_{\mathrm{ID}_I}$ to learn $\mathbf{C}$ through contrastive learning. Nevertheless, the above formulation can only optimize $\mathbf{C}$ but fail to back-propagate the gradient to update the image encoder due to the fact that the hard assignment in Eq.~(\ref{hard}) is not differentiable.  Thus, we soften the hard assignment  and map the image feature $\mathbf{x}$ into a weighted summation of codewords:
\begin{equation}
\label{soft}
\hat{\mathbf{x}} = \sum_{i=1}^N s_i \mathbf{c}_i,
\end{equation}
where $s_i$ is the weight computed by
\begin{equation}
\label{eq:soft2}
s_i = \frac{e^{-\beta \|\mathbf{x} - \mathbf{c}_i\|_2}}{\sum_{j=1}^Ne^{-\beta \|\mathbf{x} - \mathbf{c}_j\|_2}},
\end{equation}
where $\beta$ is a pre-defined positive constant controlling the softness of the assignment.  When $\beta \to +\infty$, the soft-assignment operation in Eq.~\eqref{soft} will degenerate to the hard assignment. To approximate the hard assignment well, $\beta$ cannot be set to be a small value. On the other hand, if $\beta$ is too large, the softmax function in Eq.~\eqref{soft} tends to fall into the saturation region and leads to gradient vanishing. In implementation, we set $\beta$ to be a small value in the first epoch and gradually increases it to a large value in the training process.

\begin{figure*}[t]
\centering
\includegraphics[width=7in]{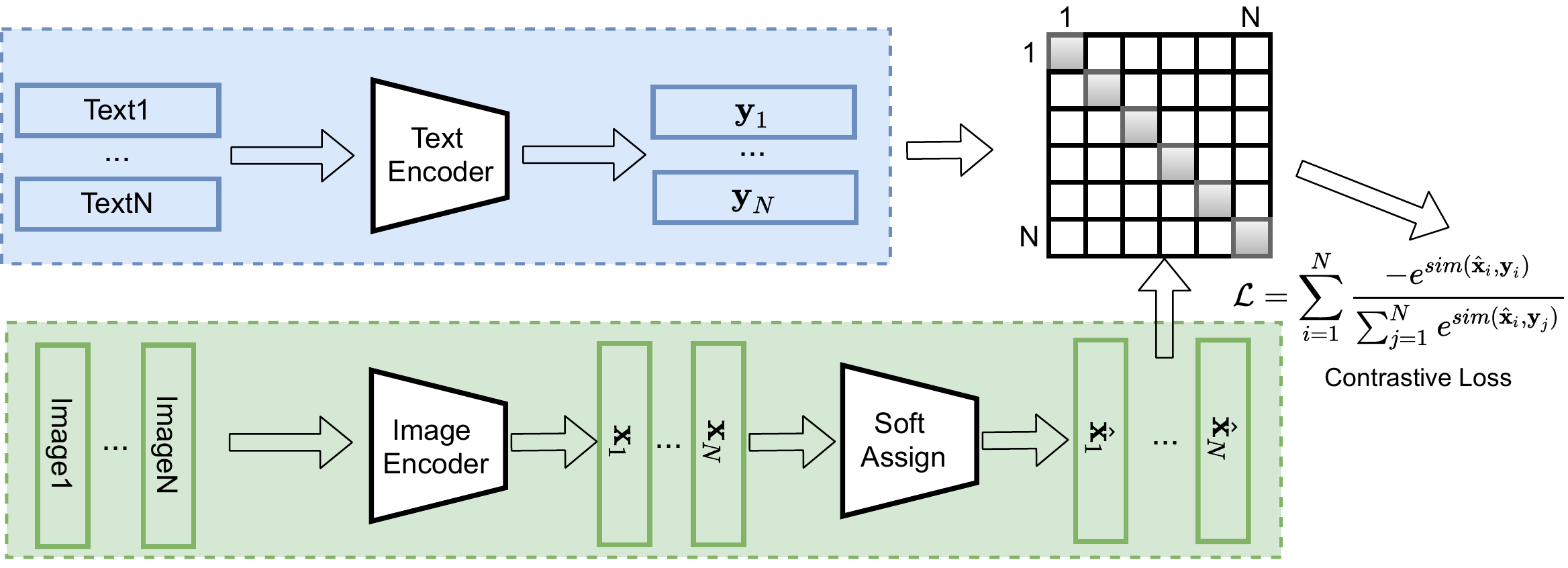}
\caption{The  pipeline of learning visual IDs.  Given a batch of $N$ text-image pairs. An image is only relevant with the text  in the pair and is irrelevant with other $N-1$ images. The text extractor generates the text features $\{\mathbf{y}_i\}_{i=1}^N$ and the image extractor generates the image features $\{\mathbf{x}_i\}_{i=1}^N$.  An image feature $\mathbf{x}_i$ is mapped into a weighted summation of its neighboring codewords $\hat{\mathbf{x}}_i$ through the soft assignment.  Then the similarity between the text features  $\{\mathbf{y}_i\}_{i=1}^N$ and the mapped image codewords  $\{\hat{\mathbf{x}}_i\}_{i=1}^N$  are optimized through the contrastive loss. }
\label{visualID}\vspace{0in}
\end{figure*}

To partition the feature space into a fine level, we need to set the number of codewords in the codebook $\mathbf{C}$ large enough. Nevertheless, the increase in the number of codewords will inevitably lead to an increase in memory and computation costs. To achieve high efficiency,  we conduct a two-level quantization by exploiting residual quantization.
In the first level, we map the feature $\mathbf{x}$ into codewords from the coarse-level dictionary $\mathcal{C}_0 = \{\mathbf{c}_{0,i}\}_{i=1}^{N_0}$ in a soft-assignment manner:
\begin{equation}
\label{eq:x}
\bar{\mathbf{x}} = \sum_{i=1}^{N_0} s_{0,i} \mathbf{c}_{0,i},
\end{equation}
where
\begin{equation}
s_{0,i} = \frac{e^{-\beta \|\mathbf{x} - \mathbf{c}_{0,i}\|_2}}{\sum_{j=1}^Ne^{-\beta \|\mathbf{x} - \mathbf{c}_{0,j}\|_2}},
\end{equation}
Then we compute the residual vector $\mathbf{r}$:
\begin{equation}
\mathbf{r} = \mathbf{x} - \hat{\mathbf{x}} .
\end{equation}
After that, we conduct  product quantization (PQ)~\cite{jegou2011product,ge2013optimized} to split the  image residual features $\mathbf{r} \in \mathbb{R}^{d}$ into $K$ segments:
\begin{equation}
\mathbf{r} \to [\mathbf{r}_1,\cdots,\mathbf{r}_K],
\end{equation}
and each segment of the residual vector $\mathbf{r}_k \in \mathbb{R}^{\frac{d}{K}}$, $\forall k \in [1,K]$.
We conduct clustering  on each split segment of the residual vector $\mathbf{r}_k$ based on the split-specific codebook $\mathcal{C}_k  = \{\mathbf{c}_{k,i}\}_{i=1}^{N_k}$, where $\mathbf{c}_{k,i} \in \mathbb{R}^{\frac{d}{K}}$.
To be specific, we conduct
\begin{equation}
\label{soft2}
\bar{\mathbf{r}}_k = \sum_{i=1}^{N_k} s_{k,i} \mathbf{c}_{k,i},
\end{equation}
where
\begin{equation}
s_{k,i} = \frac{e^{-\beta \|\mathbf{r}_k - \mathbf{c}_{k,i}\|_2}}{\sum_{j=1}^Ne^{-\beta \|\mathbf{r}_k - \mathbf{c}_{k,j}\|_2}}.
\end{equation}
After that, $\{\hat{\mathbf{r}}_k\}_{k=1}^K$  are concatenated into the recovered residual feature:
\begin{equation}
\hat{\mathbf{r}} \gets [\hat{\mathbf{r}}_1, \cdots, \hat{\mathbf{r}}_K].
\end{equation}
The recovered feature vector is obtained by summing up the  recovered residual feature $\hat{\mathbf{r}}$ and coarse-quantize vector $\bar{\mathbf{x}}$ in Eq.~(\ref{eq:x}):
\begin{equation}
\label{eq:last}
\hat{\mathbf{x}} =  \hat{\mathbf{r}} + \bar{\mathbf{x}}.
\end{equation}
Straightforwardly, the coarse-level codebook $\mathcal{C}_0$ and $K$ fine-level  sub-codebooks $\{\mathbf{C}_k\}_{k=1}^K$ can partition the whole  feature space into $\prod_{k=0}^K N_k$ cells. But the total memory complexity  of storing the  codebooks $\{\mathbf{C}_k\}_{k=0}^K$ is only $\mathcal{O}(dN)$.

\vspace{0.15in}
\noindent \textbf{Contrastive learning.} %
The contrastive learning is conducted within each mini-batch.
Given a mini-batch of text-image pairs $\{(I_i,T_i)\}_{i=1}^B$, the image $I_i$ is only relevant with the text $T_i$ and is irrelevant with other texts, $T_j$ ($j\neq i$).
Thus, after we obtain the text features $\{\mathbf{y}_i\}_{i=1}^B$ and the mapped image features $\{\hat{\mathbf{x}}_i\}_{i=1}^B$ from the codebooks $\{\mathcal{C}\}_{k=0}^K$ based on Eq.~(\ref{soft})-(\ref{eq:last}), we construct the contrastive loss defined as
\begin{align}
&\mathcal{L} =\\\notag
&- \frac{1}{B} \sum_{i=1}^B\left \{ \mathrm{log}(\frac{e^{\mathrm{sim}(\hat{\mathbf{x}}_i,\mathbf{y}_i)}}{\sum_{j=1}^B e^{\mathrm{sim}(\hat{\mathbf{x}}_j,\mathbf{y}_i)}}) + \mathrm{log}(\frac{e^{\mathrm{sim}(\hat{\mathbf{x}}_i,\mathbf{y}_i)}}{\sum_{j=1}^B e^{\mathrm{sim}(\hat{\mathbf{x}}_i,\mathbf{y}_j)}}) \right\},
\end{align}
where $\mathrm{sim}(\cdot,\cdot)$ is the function measuring the cosine similarity between two vectors defined as
\begin{equation}
\mathrm{sim}(\mathbf{x}, \mathbf{y}) = \frac{\mathbf{x}^{\top} \mathbf{y}}{\|\mathbf{x}\|_2 \|\mathbf{y}\|_2}.
\end{equation}
The contrastive learning loss aims to enlarge the gap between similarities of positive pairs and those of negative pairs. The process of learning image encoder, text encoder and dictionaries are visualized in Figure~\ref{visualID} and  summarized in Algorithm~\ref{alg:basic}.

\begin{algorithm} [h]
\caption{Image encoder and dictionary learning.}
\label{alg:basic}
    \textbf{Input}: {The image encoder $\mathrm{Encoder}_{\mathrm{img}}$,  the text encoder $\mathrm{Encoder}_{\mathrm{txt}}$, $K+1$ codebooks $\{\mathcal{C}_k\}_{k=0}^K$,  and text-image pairs $\{I_i, T_i\}_{i=1}^B$}, the positive constant $\beta$.

   \textbf{for} {$i\ \in [1,B]$}{

   ~~~$ \mathbf{x}_i \gets \mathrm{Encoder}_{\mathrm{img}}(I_i)$

   ~~~$\mathbf{y}_i \gets   \mathrm{Encoder}_{\mathrm{txt}}(T_i)$

   ~~~$\bar{\mathbf{x}}_{i} = \sum_{j=1}^N \mathbf{c}_{0,j} \frac{e^{-\beta\|\mathbf{x} - \mathbf{c}_{0,j}\|_2}}{\sum_{l=1}^N e^{-\beta\|\mathbf{x} - \mathbf{c}_{0,l}\|_2}}$

   ~~~   $ \mathbf{r}_i \gets \mathbf{x}_i - \bar{\mathbf{x}}_{i}$

   ~~~$[\mathbf{r}_{i,1}, \cdots, \mathbf{r}_{i,K}] \gets \mathbf{r}_i$\\
 \textbf{for} {$k\ \in [1,K]$}{

 ~~~$\hat{\mathbf{r}}_{i,k} = \sum_{j=1}^N \mathbf{c}_{1,j} \frac{e^{-\beta\|\mathbf{r}_{i,k} - \mathbf{c}_{k,j}\|_2}}{\sum_{l=1}^N e^{-\beta\|\mathbf{r}_{i,k} - \mathbf{c}_{k,l}\|_2}}$
 }

    ~~~$\hat{\mathbf{r}}_i \gets [\hat{\mathbf{r}}_{i,1}, \cdots, \hat{\mathbf{r}}_{i,K} ]$
    ~~~$\hat{\mathbf{x}}_i \gets \hat{\mathbf{r}}_i + \bar{\mathbf{x}}_{i}$ \\
   }
   $ \mathcal{L} = \frac{-1}{B}\sum_{i=1}^B  \mathrm{log} \left( \frac{e^{ \frac{\hat{\mathbf{x}}_i^{\top}\mathbf{y}_i} {\|\mathbf{x}_i\|_2 \|\mathbf{y}_i\|_2} }}{\sum_{j=1}^B e^{\frac{\hat{\mathbf{x}}_i^{\top}\mathbf{y}_j} {\|\mathbf{x}_i\|_2 \|\mathbf{y}_j\|_2}}} \right) $\;

   Update $\mathrm{Encoder}_{\mathrm{img}}$, $\mathrm{Encoder}_{\mathrm{txt}}$, the codebooks  $\{\mathcal{C}_k\}_{k=0}^K$ based on $\mathcal{L}$ using stochastic gradient descent.
\end{algorithm}

\begin{figure} [t!]
\includegraphics[width=3.5in]{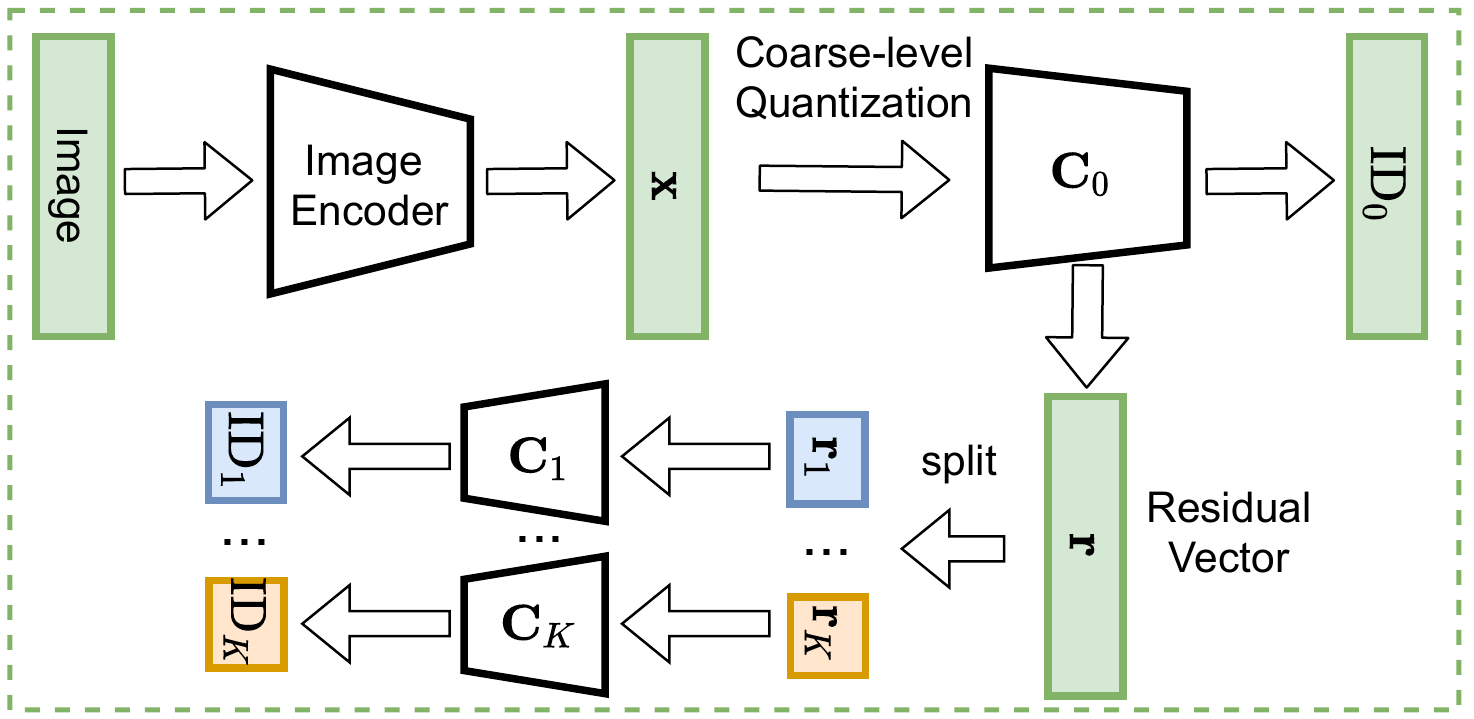}

\vspace{0.1in}

\caption{ The pipeline of generating visual IDs.  An image  is fed into the image encoder to generate the image feature $\mathbf{x}$.
It first finds its closest codeword in the  coarse-level codebook $\mathcal{C}_0 = \{\mathbf{c}_{0,i}\}_{i=1}^{N_0}$ and the index of the losest codeword  is the first visual ID, $\mathrm{ID}_0$.
Meanwhile, we obtain the residual vector $\mathbf{r} = \mathbf{x} - \mathbf{c}_{0,\mathrm{ID}_0}$, which is equally split into $K$ segments  $[\mathbf{r}_1,\cdots,\mathbf{r}_K]$.
 Each segment $\mathbf{r}_{k}$  is mapped to  its closest codeword in the dictionary $\mathbf{C}_k$ and the index of the closest codeword in $\mathbf{C}_k$ corresponds to the visual ID, $\mathrm{ID}_K$. In total, we obtain $K+1$ visual IDs for an image.}
\label{visualIDgen}
\end{figure}

\vspace{0.15in}
\noindent \textbf{Visual ID generation.} After we train the models using the contrastive learning, we use the image encoder $\mathrm{Encoder}_{\mathrm{img}}$ to extract the image feature $\mathbf{x}$. Then $\mathbf{x}$ is mapped to its closest codewords in the learned  coarse-level dictionary $\mathcal{C}_0 = \{ \mathbf{c}_{0,i}\}_{i=1}^{N_0}$. We define the index of the mapped codeword in $\mathcal{C}_0$ as the visual ID $\mathrm{ID}_0^v$. Then, we compute the residual feature vector $\mathbf{r} = \mathbf{x} - \mathbf{c}_{0,\mathrm{ID}_0^v}$.  After that, $\mathbf{r}$ is equally split into $K$ segments $\{r_k\}_{k=1}^{K}$, and each segment is   mapped to its closest codeword in the codebook  $\mathbf{C}_k = \{ \mathbf{c}_{k,i}\}_{i=1}^{N_k}$. We define  the index of the closest codeword  in each dictionary $\mathbf{C}_k$ as the visual ID $\mathrm{ID}_k^v$, that is,
\begin{equation}
\begin{split}
\mathrm{ID}_{k}^v = \mathrm{argmin}_{i}{\|\mathbf{r}_k - \mathbf{c}_{k,i}\|}. \\
\end{split}
\end{equation}
We visualize the process of generating visual ID in Figure~\ref{visualIDgen}.
In total, for each image, we generate $K+1$ visual IDs ($\{\mathrm{ID}_{k}^v\}_{k=0}^K$), which will be used  as the input of the CTR model to encode the visual content of the ad.  It is worth noting that, the image extractor $\mathrm{Encoder}_{\mathrm{img}}$  and $K+1$ codebooks $\{\mathcal{C}_k\}_{k=0}^K$  are only used in generating visual ID. They will not be involved when learning the visual ID embedding  for CTR prediction.  An alternative choice is to plug the image extractor $\mathrm{Encoder}_{\mathrm{img}}$  and the codebooks $\{\mathcal{C}_k\}_{k=0}^K$ into the CTR prediction model to achieve end-to-end training. But it is prohibitively expensive due to the fact that the image feature extractor is normally heavy.

\section{CTR Prediction Model Learning}
\vspace{0.08in}
The CTR prediction model  in our advertising platform takes the user ID, the ad ID and the ad  visual ID as input to predict the CTR. We denote a user by $u$ and an ad by $a$.
The CTR prediction model consists of  embedding layers,  the ad  tower, the user  tower and the prediction head.
To be specific, the user embedding layer maps the ID of the user $u$ to a vector:
\begin{equation}
\mathbf{u} = \mathrm{EMB}_{\mathrm{user}}(\mathrm{ID}_u),
\end{equation}
the ad embedding layer maps the  ID  of the ad $a$ to a vector:
\begin{equation}
\mathbf{a} = \mathrm{EMB}_{\mathrm{ad}}(\mathrm{ID}_a).
\end{equation}
The visual embedding layer maps  the ad's each  visual IDs  $\mathrm{ID}^v_k$ ($k\in[0,K]$)  to  a feature vector in the following manner
 \begin{equation}
\begin{split}
{\mathbf{x}} &=   \mathrm{EMB}_{\mathrm{visual}}^{(0)}(\mathrm{ID}^v_{0})\\
\mathbf{r}_k &= \mathrm{EMB}_{\mathrm{visual}}^{(k)}(\mathrm{ID}^v_{k}),~\forall k\in[1,K]. \\
\end{split}
\end{equation}
Then the visual embedding of the ad $\mathbf{v}$ is obtained by concatenating $\{\mathbf{r}_k\}$ and summing up the concatenated vector with ${\mathbf{x}}$
\begin{equation}
\begin{split}
\mathbf{r} &=  [\mathbf{r}_1,\cdots, \mathbf{r}_K],\\
\mathbf{v} &= \mathbf{x} + \mathbf{r} .\\
\end{split}
\end{equation}
The user  tower takes $\mathbf{u}$ as input, and generates the user feature:
\begin{equation}
\hat{\mathbf{u}} = \mathrm{Tower}_{\mathrm{user}}(\mathbf{u}).
\end{equation}
In parallel, the ad tower takes $\mathbf{a}$ and $\mathbf{v}$ as input, and generates the ad feature:
\begin{equation}
\hat{\mathbf{a}} = \mathrm{Tower}_{\mathrm{ad}}([\mathbf{a},\mathbf{v}]).
\end{equation}
In practice, both $\mathrm{Tower}_{\mathrm{ad}}$ and $ \mathrm{Tower}_{\mathrm{user}}$ are implemented by  several fully-connected layers.
The prediction head takes the user feature $\hat{\mathbf{u}} \in \mathbb{R}^D $ and the ad feature $\hat{\mathbf{a}} \in \mathbb{R}^D$ as input and generates the predicted CTR:
\begin{equation}
\hat{y} = \mathrm{sigmoid}([\hat{\mathbf{u}},\hat{\mathbf{a}}] \mathbf{w} + b),
\end{equation}
where $\mathbf{w} \in \mathbb{R}^{2D}$ and $b \in \mathbb{R}$ are weights of the prediction head, and $\mathrm{sigmoid}$ is the function defined as $\mathrm{sigmoid}(x) = \frac{1}{1+e^{-x}}$.
In the training phase, we construct a binary cross-entropy loss to update the weights CTR prediction model.  To be specific, given the predicted CTR  $\hat{y} \in (0,1)$ and  the ground-truth CTR by $y \in \{0,1\}$, the binary cross-entropy loss is constructed by
\begin{equation}
\mathcal{L}_{\mathrm{BCE}} = -[(1-y)\mathrm{log}(\hat{y}) + y\mathrm{log}(1-\hat{y})].
\end{equation}

\begin{algorithm}[h]
\caption{CTR prediction model learning.}
\label{alg:basic2}
    \textbf{Input}: {The user-ad pairs $\{a_i,u_i\}_{i=1}^M$ and the ground-truth labels $\{y_i\}_{i=1}^M$, user embedding layer $\mathrm{EMB}_{\mathrm{user}}$, ad embedding layer $\mathrm{EMB}_{\mathrm{ad}}$, and  visual embedding layers  $\mathrm{EMB}_{\mathrm{visual}}^{(1)}$ and  $\mathrm{EMB}_{\mathrm{visual}}^{(2)}$, user tower $\mathrm{Tower}_{\mathrm{user}}$, ad tower $\mathrm{Tower}_{\mathrm{ad}}$, the prediction head with weights $\{\mathbf{w},b\}$}\;

   \textbf{for}~~{$i\ \in [1,M]$}{

   ~~~fetch the ad ID for $a_i$, $\mathrm{ID}_{a}$\;

   ~~~$\mathbf{a} = \mathrm{EMB}_{\mathrm{ad}}(\mathrm{ID}_a)$ \;

   ~~~fetch the visual IDs of $a_i$, $\{\mathrm{ID}^v_{k}\}_{k=0}^K$\;

   ~~~$\mathbf{x} = \mathrm{EMB}_{\mathrm{visual}}^{(0)}(\mathrm{ID}_{0}^v)$\;

      \textbf{for}~~{$k\ \in [1,K]$}{

   ~~~$\mathbf{r}_k = \mathrm{EMB}_{\mathrm{visual}}^{(k)}(\mathrm{ID}_{k}^v)$\;
   }

   ~~~$\mathbf{r} = [\mathbf{r}_1,\cdots, \mathbf{r}_K]$\;

    ~~~$\mathbf{v} = \mathbf{r} + \mathbf{x}$\;

   ~~~$\hat{\mathbf{a}} = \mathrm{Tower}_{\mathrm{ad}}([\mathbf{a},\mathbf{v}]$ \;

  ~~fetch the user ID for $u_i$, $\mathrm{ID}_{u}$\;

   ~~~$\mathbf{u} = \mathrm{EMB}_{\mathrm{user}}(\mathrm{ID}_u)$\;

   ~~~$\hat{\mathbf{u}} = \mathrm{Tower}_{\mathrm{user}}(\mathbf{u})$\;

   ~~~$\hat{y}_i = \mathrm{sigmoid}([\hat{\mathbf{u}},\hat{\mathbf{a}}] \mathbf{w} + b),$
   }

   $ \mathcal{L} = - \frac{1}{M}\sum_{i=1}^M (1-y_i)\mathrm{log}(\hat{y}_i) + y_i\mathrm{log}(1-\hat{y}_i) $\;

   Update $\mathrm{EMB}_{\mathrm{user}}$,  $\mathrm{EMB}_{\mathrm{ad}}$,    $\mathrm{EMB}_{\mathrm{visual}}^{(1)}$,  $\mathrm{EMB}_{\mathrm{visual}}^{(2)}$,  $\mathrm{Tower}_{\mathrm{user}}$,  $\mathrm{Tower}_{\mathrm{ad}}$, $\mathbf{w}$, and $b$ using $\mathcal{L}$ by stochastic gradient descent.
\end{algorithm}

\noindent Algorithm~\ref{alg:basic2} summarizes the CTR prediction model learning.

 \vspace{0.1in}

 \section{Experiments}
 \vspace{0.1in}

\noindent\textbf{Implementation.} We implement the image encoder $\mathrm{Encoder}_{\mathrm{img}}$ by ResNet50~\cite{he2016deep} with $50$ layers  and the text encoder   $\mathrm{Encoder}_{\mathrm{txt}}$ by the BERT-base model~\cite{devlin2019bert}  with $12$ Transformer layers and $768$ hidden size.  ResNet50 is pre-trained on ImageNet dataset and the BERT-base model is pre-trained on sentences from the BooksCorpus dataset with 800 Million words and English Wikipedia with 2,500 million words. We use the hidden feature before the last classification layer of ResNet50 as the image feature $\mathbf{x}$ and use the hidden state of the [cls] token in the last layer of the BERT-base as the text feature $\mathbf{y}$.    To make the dimension of the image feature identical to that of the text feature, we add a fully-connected layer to project the feature obtained from the ResNet to  the $768$-dimensional vector.
By default, we set the number of codebooks for the residual vector, $K$, as $4$.  Meanwhile we set the default number of codewords in each codebook $\mathcal{C}_k$, $N_k$, as $256$. In the training process, we set the initial value of $\beta$ as $1$ and gradually increase it to $10$.

\vspace{0.08in}
\noindent \textbf{Training data.} To train the image extractor  $\mathrm{Encoder}_{\mathrm{img}}$, the text extractor  $\mathrm{Encoder}_{\mathrm{txt}}$ and the codebooks $\{\mathcal{C}_k\}_{k=0}^K$, we use $50$ million text-image pairs collected from our advertising system.  To train the CTR prediction model, we  use $70$ million user-click  data per day collected from our advertising system and the model is fine-tuned for around 1 month.

\subsection{Offline Experiments}

We evaluate  the influence of our model on the CTR prediction in the offline phase. We first define the CTR prediction model using only ad behavior ID features and user behavior ID features as the baseline. We use the AUC improvement of CTR prediction over the baseline model as the evaluation metrics. Meanwhile, we use an observation window of 40-day length to demonstrate more details.

\begin{figure}[h]
\vspace{-0.1in}

 \centering
    \includegraphics[width=2.65in]{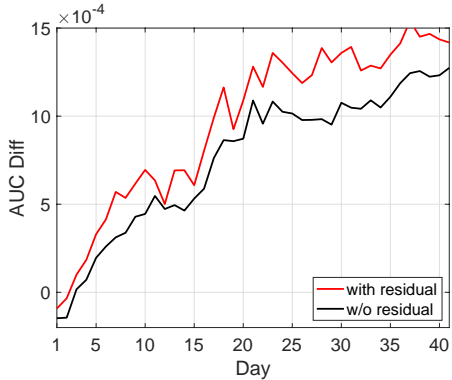}

\vspace{-0.1in}

     \caption{Comparisons with the setting without residual quantization.}
     \label{fig:res}
 \end{figure}

\vspace{0.1in}
\noindent \textbf{Residual quantization.}  By default, we conduct a two-level residual quantization. In the first level, we conduct a coarse level quantization using $\mathcal{C}_0$ and then conduct the second-level quantization on the residual. Here, we compare the two-level residual quantization with a single-level quantization by removing the first-level coarse level quantization.
We report the AUC difference between the model with the baseline model along $40$ days in Figure~\ref{fig:res}.

As shown in the figure, using the residual quantization, the model can get more improvement compared with its counterpart without residual quantization.  To be specific, at the $40$-th day, using residual quantization, the model achieves a $0.1418\%$ AUC improvement, whereas the model without residual quantization only achieves a $0.1242\%$ AUC increase.





\begin{figure}[t]
\centering
    \includegraphics[width=2.8in]{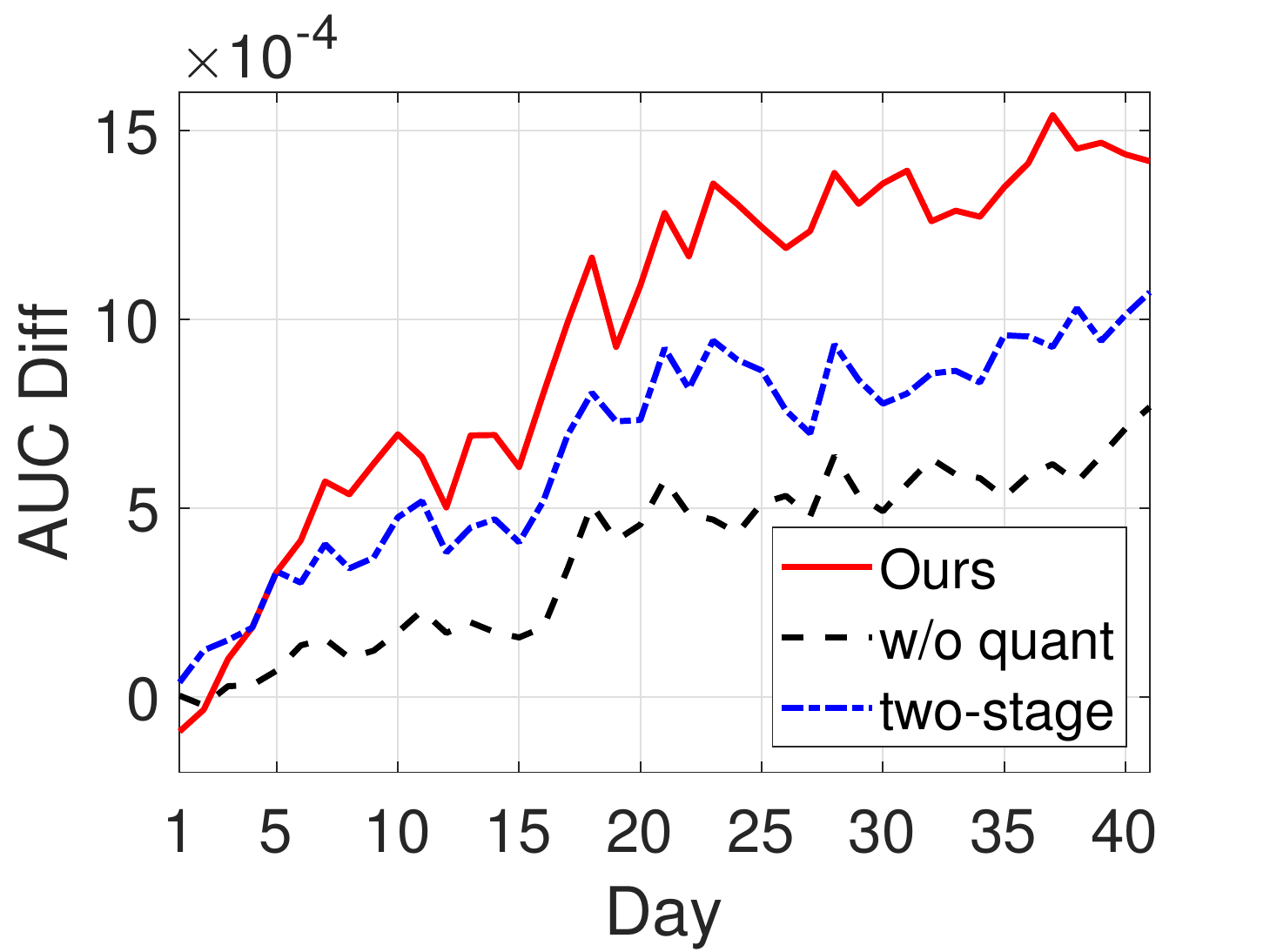}

\vspace{-0.1in}

     \caption{Comparisons with the two-stage setting and the setting without quantization.}
     \label{fig:quan}\vspace{-0.2in}
 \end{figure}

\vspace{0.1in} \noindent \textbf{Comparisons with alternative configurations.} We soften the quantization operation as Eq.~\eqref{soft} and Eq.~\eqref{eq:soft2} to make it differentiable. Thus, the codeword assignment can be incorporated into a neural network, and the codewords can be learned in an end-to-end manner.  An alternative solution is a two-stage learning process. In the first stage, we can learn the raw feature vectors without quantization  through the contrastive loss based on the text-image pairs.  After that, we use the unsupervised k-means clustering to conduct residual quantization and product quantization for generating the codebooks and the visual IDs.  We term this configuration as the two-stage baseline.  Meanwhile, we also compare with the baseline directly using the raw  feature vectors learned  without quantization  through the contrastive loss based on text-image pairs   as the visual representation for CTR prediction. We term it as w/o quantization baseline. As shown in Figure~\ref{fig:quan}, ours consistently outperforms two compared baselines, demonstrating the effectiveness of learning  codebooks  in an end-to-end manner.

 \begin{figure}[b!]

\vspace{-0.25in}
\centering

     \subfigure[The influence of $N$]{\includegraphics[width=2.65in]{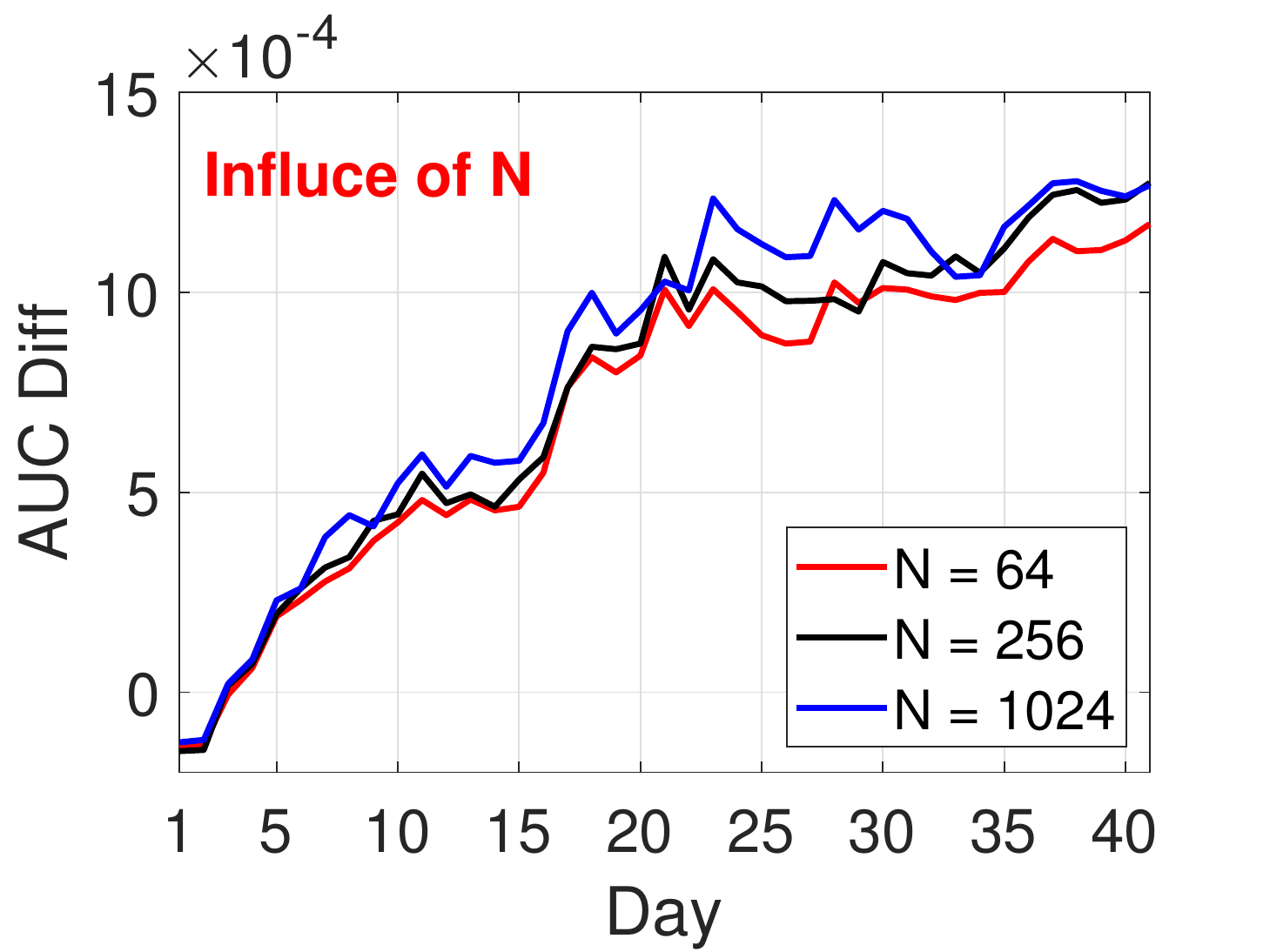}}\\

\vspace{-0.1in}

      \subfigure[The influence of $K$]{\includegraphics[width=2.65in]{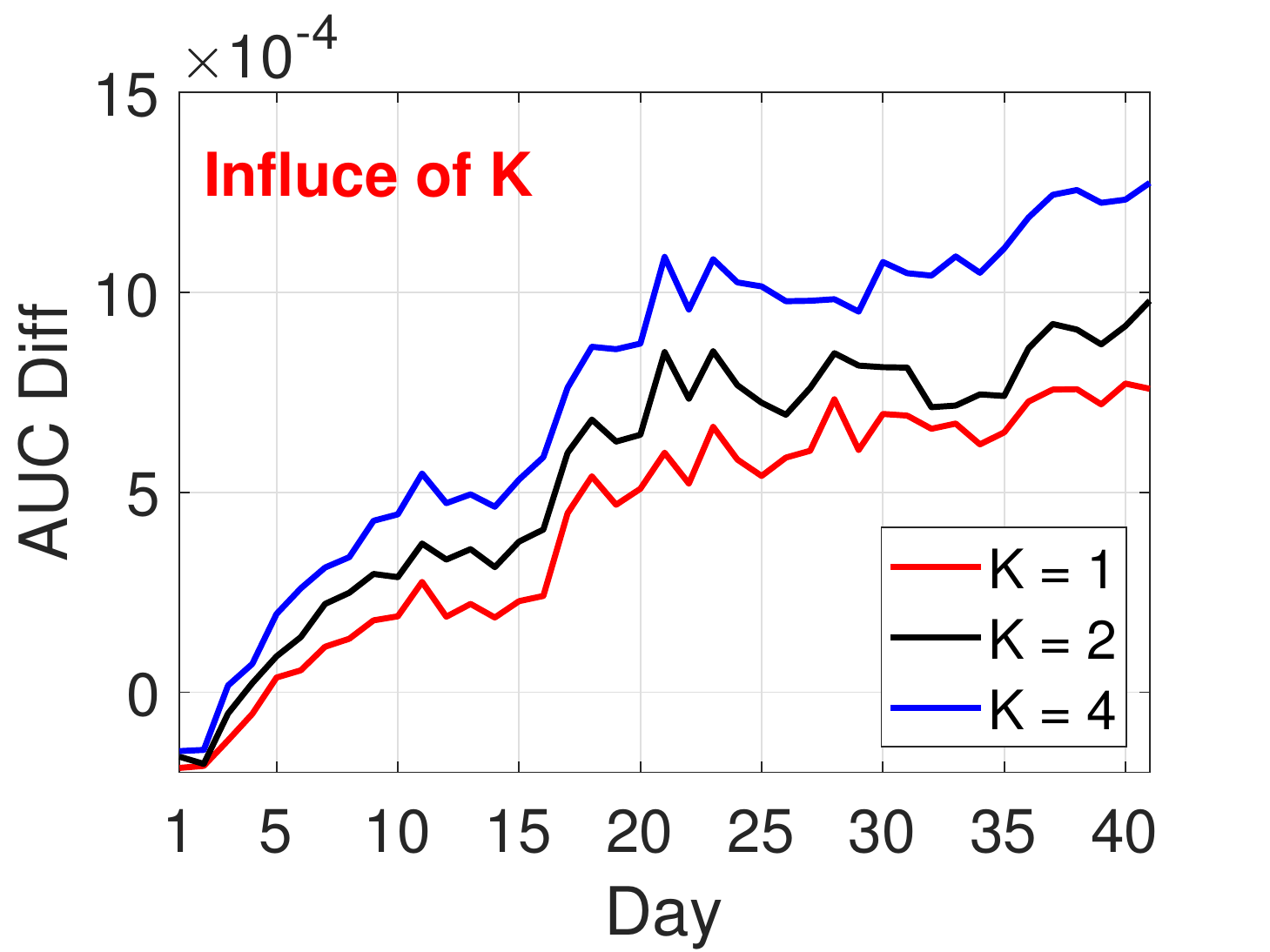}}

\vspace{-0.1in}

     \caption{The influence of parameters on the performance.}
     \label{fig:hyper}\vspace{-0.1in}
 \end{figure}

 \begin{figure*}[t]
    \centering
     \subfigure[$\mathrm{ID}^v_0=1388$, $\mathrm{ID}^v_1=145$, $\mathrm{ID}^v_2=251$, $\mathrm{ID}^v_3=86$, $\mathrm{ID}^v_4=87$]{
        \includegraphics[width=1\textwidth, height=0.17\textwidth ]{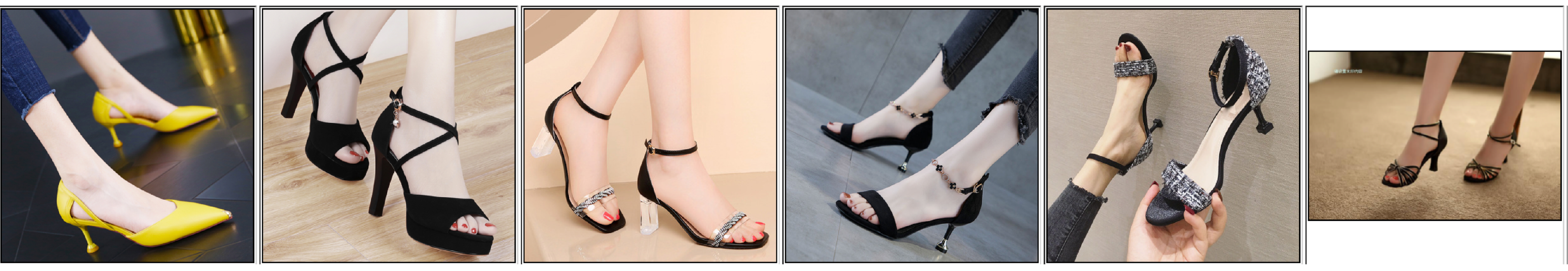}}
    \subfigure[$\mathrm{ID}^v_0=1511$, $\mathrm{ID}^v_1=201$, $\mathrm{ID}^v_2=106$, $\mathrm{ID}^v_3=42$, $\mathrm{ID}^v_4=149$]{
        \includegraphics[width=1\textwidth, height=0.17\textwidth]{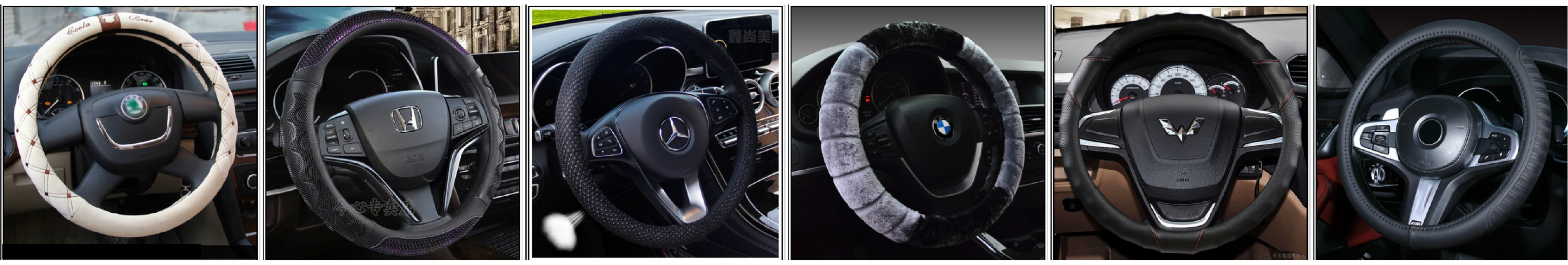}}
    \subfigure[$\mathrm{ID}^v_0=84$, $\mathrm{ID}^v_1=104$, $\mathrm{ID}^v_2=61$, $\mathrm{ID}^v_3=22$, $\mathrm{ID}^v_4=202$]{
        \includegraphics[width=1\textwidth, height=0.17\textwidth]{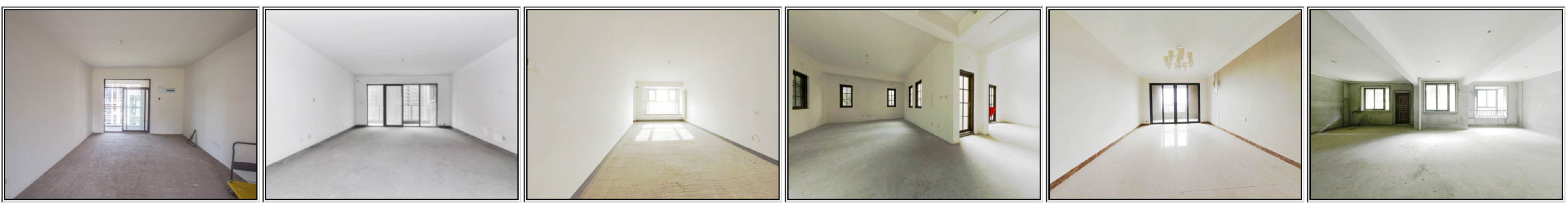}}
       \caption{Visualization of ads with the same visual IDs.}
    \label{fig:visual}
\end{figure*}

\vspace{0.05in}
\noindent \textbf{Influence of the number of codewords.} A larger number of codewords can partition the feature space into finer cells and leads to smaller quantization errors.
We evaluate the influence of the number of codewords, $N$, on the AUC of CTR prediction.   To this end, we show the results when $N$ varies among $\{64,256,1024\}$. As shown in Figure~\ref{fig:hyper} (a), when $N$ increases from $64$ to $256$, the AUC of CTR prediction gets improved considerably.   But when $N$ further increases to $1024$, the AUC gets saturated. Since the increase of $N$ will bring more computation cost, considering both efficiency and effectiveness, we set $N=256$ by default.

\vspace{0.1in}
\noindent \textbf{Influence of the number of segments.}  We split the residual vector into $K$ segments and conduct product quantization based on split segments. A larger $K$ also leads to a
finer partition of the feature space and smaller quantization errors.  We evaluate the influence of the number of segments, $K$, on the AUC of CTR prediction. To this end, we show the results when $K$ is chosen from $\{1,2,4\}$.   As shown in Figure~\ref{fig:hyper} (b), when $K$ increases from $1$ to $4$, the AUC of CTR prediction  gets boosted significantly.
To be specific, when $K=1$, at the $40$-th day, the AUC improvement over baseline is only $0.0903\%$, whereas the  AUC improvement is $0.1418\%$ when $K=4$.
 But the increase of $K$ also brings more computational cost. Thus, to balance the efficiency and effectiveness, we do not further increase $K$ beyond $4$ and  set $K=4$, by default.


\vspace{0.1in}
\noindent \textbf{Influence on new ads and old ads.} As we know, incorporating the visual content cannot bring more benefits to new ads since their ID features are not reliable. Here, we demonstrate the influence of our model in new ads and the old ones.  As shown in Table~\ref{tab:newold}, using the visual ID embedding, it has a more influential impact on new ads than old ads.
To be specific,  for new ads, it brings $+0.6098\%$  increase in AUC of the CTR prediction, whereas it only brings $+0.1238\%$ increase for old ads.

\begin{table}[t]


\caption{The influence of visual ID embedding on new ads and old ads.}
\centering
\begin{tabular}{cc|cc} \hline
\multicolumn{2}{c|}{old ads} & \multicolumn{2}{c}{new ads} \\ \hline
\# of instances  & AUC diff & \# of instances  &  AUC diff \\ \hline
          $61,982,597$       &   $+0.1238\%$   &       $4,935,208$           &  $+0.6098\%$   \\\hline
\end{tabular}
\label{tab:newold}\vspace{-0.1in}
\end{table}

\vspace{0.1in}
\noindent \textbf{Visualization.} In Figure~\ref{fig:visual}, we visualize the ads with the same visual IDs, $\{\mathrm{ID}_{k}^v\}_{k=0}^4$. It is not difficult to observe from the figure that the ads with the same visual IDs have a similar visual appearance, demonstrating the effectiveness of the learned visual IDs.

\subsection{Online Experiments}

  \begin{table}[h]
\centering

  \caption{The influence of launching the proposed model in our advertising platform. Note that ``$+\cdot\%$'' is  relative improvement for online experiments.}
\begin{tabular}{cc}\\ \hline
  Charge & CTR \\ \hline
 $+1.10\%$ & $+1.46\%$\\ \hline
\end{tabular}
\label{online}
\end{table}

In Table~\ref{online}, we show the influence of launching the proposed model in Baidu online advertising platform. Before launching our model, the CTR prediction model in our advertising platform uses only the ad ID features and user ID features based on the users' historical behaviors.  As shown in the table, after launching our model,  the charge of our advertising platform gets increased relatively by $+1.10\%$, and the CTR of the ads gets improved relatively by $+1.46\%$.  That is, the launching of the proposed model brings significant revenue boosts for the advertisers as well as our ads platform.

\vspace{0.1in}

\section{Conclusion}

\vspace{0.1in}

The traditional CTR prediction model relies on the user behavior ID features and product (ad) behavior ID features. But the effectiveness of these ID features requires rich historical user-product (ad) behaviors, and thus they do not generalize well to the new products (ads). In this work, we investigate  exploiting visual content in ads to boost the CTR prediction accuracy. We adopt the textual description of images as the supervision to learn the image feature extractor and map the visual content in each ad to visual IDs.   Then we learn the embedding  for visual IDs based on the user's clicking behaviors. We feed the learned visual ID embedding besides the ad behavior ID embedding and the user  behavior ID embedding  into a CTR prediction model to generate a more reliable CTR prediction. The  systematic offline experiments demonstrate the improved AUC of the  CTR prediction  by incorporating the visual content. After launching the proposed model in Baidu advertising platform, we achieve a $ 1.10 \%$ charge increase  and a $1.46\%$ CTR increase.

\vspace{0.5in}

\bibliographystyle{plain}
\bibliography{refs_scholar}

\end{document}